# High mobility conduction at (110) and (111) LaAlO$_3$/SrTiO$_3$ interfaces

G. Herranz[*], F. Sánchez, N. Dix, M. Scigaj, J. Fontcuberta

Institut de Ciència de Materials de Barcelona (ICMAB-CSIC), Campus de la UAB, Bellaterra 08193, Catalonia, Spain

Email: gherranz@icmab.es

## Abstract

**In recent years, striking discoveries have revealed that two-dimensional electron liquids (2DEL) confined at the interface between oxide band-insulators can be engineered to display a high mobility transport. The recognition that only few interfaces appear to suit hosting 2DEL is intriguing and challenges the understanding of these emerging properties not existing in bulk. Indeed, only the neutral TiO$_2$ surface of (001)SrTiO$_3$ has been shown to sustain 2DEL. We show that this restriction can be surpassed: (110) and (111) surfaces of SrTiO$_3$ interfaced with epitaxial LaAlO$_3$ layers, above a critical thickness, display 2DEL transport with mobilities similar to those of (001)SrTiO$_3$. Moreover we show that epitaxial interfaces are not a prerequisite: conducting (110) interfaces with amorphous LaAlO$_3$ and other oxides can also be prepared. These findings open a new perspective both for materials research and for elucidating the ultimate microscopic mechanism of carrier doping.**

High mobility conduction at (110) and (111) LaAlO$_3$/SrTiO$_3$ interfaces

**Introduction**

Few years ago it was discovered that the interface between the band insulators LaAlO$_3$ (LAO) and SrTiO$_3$ (STO) can be metallic with electron mobilities up to a few thousands of cm$^2$/Vs and sheet carrier densities slightly above 10$^{13}$ cm$^{-2}$ [1]. Quite remarkably, such conductance is extremely confined, within a few unit cells from the interface [2, 3]. Following this hallmark, it has been shown that this two-dimensional electron liquid (2DEL) can host superconducting condensates[2] or even incipient magnetism [4, 5], illustrating how electron correlations can be enhanced in 2D confined systems thus suggesting new avenues for engineering emerging properties at interfaces. In spite of the enormous progress made on tailoring new properties [6, 7, 8] or designing devices [9, 10], the scarcity of materials and interfaces displaying 2DEL behavior remains an exciting open question. Indeed, 2DEL behavior has only been deeply investigated and reported in (001)STO interfaces along with some other few oxides (i.e., LaGaO$_3$ [11] and LnTiO$_3$, where Ln is a rare earth [12]). The scarcity of explored alternative approaches could be related to the distinct emphasis given to the different scenarios explaining the 2D conductivity, namely, oxygen vacancy doping [13, 14, 15], cation interdiffusion [16, 17] and polar discontinuity across the LAO/STO interface [1, 18]. Whereas the former two pictures rely on subtle chemical substitutions that dope the interface with carriers, the latter mechanism is purely electrostatic and is based on charge transfer triggered by the stacking of charged ionic LAO planes over the neutral STO surfaces in (001)-oriented heterostructures. The relatively easy preparation of appropriate (001)STO surfaces for heteroepitaxial growth and the beauty of the polar catastrophe picture and its ability to naturally explain some characteristic features (such as the existence critical thickness for 2DEL behavior) have probably biased research in its favor and thus the polar discontinuity scenario gained a wider acceptance. The observed transition from

High mobility conduction at (110) and (111) LaAlO$_3$/SrTiO$_3$ interfaces

insulator to metallic interfaces above a critical LAO thickness (4 monolayers (MLs) for (001)-interfaces) has been used as a strong argument to support this picture [19]. Note that the nature of the charge transfer should be strongly dependent on the interface orientation. In particular, in the ideal ionic limit, (001)- and (111)-oriented interfaces exhibit polar discontinuity because of the alternate stacking of [LaO]$^+$[AlO$_2$]$^-$ planes over [SrO]$^0$ [TiO$_2$]$^0$, or [LaO$_3$]$^{3-}$ [Al]$^{3+}$ over [SrO$_3$]$^{4-}$ [Ti]$^{4+}$, respectively (Figure 1 and Supplementary Table S1) [20]. On the contrary, the ABO$_3$ perovskite structures stack according to the sequence [ABO]$^{4+}$/[O$_2$]$^{4-}$ along [110] whatever the composition of the perovskite and, therefore, (110)-oriented LAO/STO interfaces do not exhibit any polar discontinuity. Thus, it has been generally accepted that (110)-interfaces should not promote any electronic reconstruction and therefore no interface metallicity is expected. This was found to be compatible with the observed insulating behavior for the few (110)-oriented interfaces reported to date (see, e.g., Ref. [21, 22]).

Aiming to broaden the scope of interfaces displaying 2DEL behavior, we have undertaken a systematic exploration of interfaces based on the growth of LAO along distinct atomic planes, namely on (110)STO and (111)STO, as well as by growing amorphous overlayers of different oxides, thus releasing the constrain of epitaxial interface matching (see sketches in Figure 1). It will be shown that LAO/STO interfaces with (110)- and (111)- orientations can also be made conductive, with values of the sheet carrier density and electronic mobility (up to $\approx 1 \times 10^{14}$ cm$^{-2}$ and $\approx 2500$ cm$^2$/Vs, respectively) very similar to those found in (001)-interfaces [1, 2, 3, 15, 16]. An abrupt jump from an insulator to a metallic state is also observed above critical thickness 7 MLs and 9 MLs for (110) and (111) interfaces, respectively -1 ML $\approx 2.68$ Å for (110) and 1 ML $\approx 2.19$ Å for (111), assuming a pseudocubic unit cell for LAO-. Our results, while

High mobility conduction at (110) and (111) LaAlO$_3$/SrTiO$_3$ interfaces

pointing that a crude polar discontinuity scenario is insufficient to account for the metallic state at the LAO/STO interface, definitely open up new opportunities for interfacing STO with other oxides. Supporting this view, we discovered that metallic behavior at (110) and (111) interfaces can also be obtained by using amorphous LAO, STO and yttria-stabilized zirconia oxides (YSZ).

**Results**

LAO/STO interfaces were prepared by pulsed laser deposition (see Methods) with LAO overlayers of different thickness (from 0 to 36 LAO(110) MLs and 44 LAO(111) MLs) on thermally treated (110) STO and (111) STO single crystals. The thermally treated STO substrates presented atomically flat terraces with steps corresponding to the (110) or (111) STO interplanar distance in height (Supplementary Figure S1). The reflection high energy electron diffraction (RHEED) pattern of the (110) STO substrate prior to starting LAO deposition, taken along [001] STO direction at 850 ºC and 10$^{-4}$ mbar, shows a high intensity specular spot, Bragg spots and Kikuchi lines (Figure 2a). The intensity of the specular spot was monitored during subsequent LAO growth (Figure 2b). The upper curve in Figure 2b was recorded during deposition of the 36 (110) LAO MLs (the first 14 oscillations are shown). The second and third oscillations are of low amplitude, but the next ones recover strongly in amplitude signalling layer-by-layer growth. The oscillations maintain the amplitude up to around 10 MLs, then there is a progressive damping and finally the oscillations vanish. The persistence of the oscillations before they vanished allows determining in-situ and very accurately the growth rate (30 laser pulses/monolayer in this case) and thus the deposition can be stopped at preselected thickness at wish. For 36MLs, ex-situ thickness measurement by X-ray reflectometry (Supplementary Figure S2) indicated a value of 96 +/- 5 Å, in

High mobility conduction at (110) and (111) LaAlO$_3$/SrTiO$_3$ interfaces

agreement with the corresponding number of RHEED oscillations. The RHEED pattern at the end of the deposition (36 MLs) was streaky (Figure 2c), pointing to a flat surface. Similarly, films of intermediate thicknesses were grown: Figure 2b (middle) shows the oscillations corresponding to the 10 MLs (110) LAO film, with the arrow marking the end of the deposition. Finally we note that RHEED oscillations can be clearly appreciated even for the thinnest films, as illustrated in Figure 2b (bottom) for the 4 MLs film. AFM topographic images of the 4, 10 and 36 MLs films are presented in Figures 2d-f, respectively. The three samples show morphology of terraces and steps 1 ML high, with low root-means-square roughness of 1.0-1.3 Å. Differences in morphology are caused by the different miscut of the substrates, with average terrace width of 375, 140, and ~ 100 nm for the 4, 10 and 36 MLs films. Respect to the (111) LAO films, we did not monitor intensity oscillations as depositions were done simultaneously on (110) and (111) substrates, but their streaky RHEED patterns and terraces and steps morphology (Supplementary Figure S3) suggest similar layer-by-layer growth mechanism.

In Figure 3 we depict the dependence of the room-temperature conductance of (110)- and (111)- interfaces as a function of LAO thickness ($t$). The first observation is that the resistance $R_{xx}$ is strongly dependent on the LAO thickness. Whereas for $t < 7$ (110) MLs (9 (111) MLs), the $R_{xx}$ values indicate an insulating or poorly conductive behavior (> 10 MΩ), an abrupt transition to metallicity was observed at 7 MLs and 9 MLs for (110)- and (111)- interfaces, respectively, with conductances ($G_{xx} = 1/R_{xx}$) at room temperature of $G_{xx} \approx 6.5 \times 10^{-5}$ Ω$^{-1}$ (110) and $G_{xx} \approx 2 \times 10^{-5}$ Ω$^{-1}$ (111). In order to better compare these properties with those usually reported on (001)LAO/STO interfaces, in Figure 3 we include also the room-temperature conductance as a function of the LAO overlayer

High mobility conduction at (110) and (111) LaAlO$_3$/SrTiO$_3$ interfaces

thickness reported in the literature for (001)- samples [19, 23]. We emphasize that, above the metal-to-insulator transition, the conductance of all the interfaces is very similar, whatever their crystallographic orientation. Another important observation is that while conventional (001)- interfaces exhibit an insulator-to-metal transition at a threshold thickness ($t_c$) about 4 MLs, for (110)- and (111)- interfaces $t_c$ occurs at ≈ 7 and 9 MLs, respectively (Figure 3). We note that bare STO substrates treated during 10$^3$ seconds (equivalent to the time required to grow 33 (110) MLs of LAO) in the same conditions of temperature and oxygen pressure and including post-growth annealing in oxygen rich atmosphere as used for the LAO/STO growth (see Methods) do not exhibit any measurable conductance above the experimental limit (~ 1 nΩ$^{-1}$). Additionally, as shown in Supplementary Figure S4, we have checked the robustness of the conductive state against different substrate thermal treatments that are susceptible to trigger surface reconstructions or chemical rearrangements on the STO surface state [24].

As shown in Figure 4, it is observed that for both interface orientations and for $t > t_c$, the temperature-dependent resistivity is strongly dependent on the LAO thickness. For (110) interfaces (Figure 4a), the resistance for $t$ = 8 MLs is $R_{xx}$ ≈ 15 kΩ and decreases by two orders of magnitude at T = 5 K ($R_{xx}$ ≈ 160 Ω), displaying a metallic behavior for all the range of temperatures down to 5 K. Above 10 MLs, the resistance becomes progressively larger for increasing LAO thickness and a low-temperature resistivity upturn, indicating electronic localization, starts to appear above 18 MLs. For instance, at room temperature, the resistance of the 36 MLs interface is $R_{xx}$ ≈ 1.1 × 10$^5$ kΩ and at T = 5 K it is $R_{xx}$ ≈ 4.6 × 10$^6$ kΩ, displaying resistance values about 1 to 4 orders of magnitude larger than those of the 8 MLs interface in all the temperature range (Figure 4a). On the other hand, the (111)-interfaces start to conduct at 9 MLs, with room

High mobility conduction at (110) and (111) LaAlO$_3$/SrTiO$_3$ interfaces

temperature resistance $R_{xx} \approx 7.1$ kΩ and $R_{xx} \approx 195$ Ω at T = 5K (see Figure 4b). In this case, the transport properties also evolve rapidly with thickness, and above 10 MLs the interfaces become poorly conductive. Compared to the (110)- interfaces, the transport in the (111)- counterparts degrade significantly faster with the LAO thickness. For instance, whereas the (110)-oriented 18 MLs interface exhibits a resistance $R_{xx} \approx 120$ kΩ at room temperature, the corresponding (111)-interface (*t* = 22 MLs) has a resistance $R_{xx} \approx 5.6$ MΩ (Figure 4b).

To have a better comprehension of the transport mechanisms, we will focus now on the evolution of the sheet carrier density and mobility of the (110)-oriented LAO/STO interfaces displayed in Figure 5. Above the insulator-to-metal transition (> 7 MLs), the sheet carrier density at low temperature is $n_{sheet}$ (5K) $\approx 10^{13} - 10^{14}$ cm$^{-2}$, comparable to the values reported for (001)-oriented surfaces (Figure 5a) [1, 2, 3, 15, 17, 23]. Note that the sheet carrier density increases slightly with the temperature; a similar variation is also observed in (001)-oriented interfaces (see e.g. Ref. [23]). On the other hand, the electron mobility displays a strong dependence on the LAO overlayer thickness (Figure 5b). For 8 MLs, the mobility at low temperature is $\mu_{5K} \approx 680$ cm$^2$/Vs, and it subsequently increases to $\mu_{5K} \approx 2500$ cm$^2$/Vs for 10 MLs. For larger thickness the mobility decreases significantly, displaying values $\mu_{5K} \approx 880$ cm$^2$/Vs for 14 MLs, and $\mu_{5K} \approx 30$ cm$^2$/Vs for 18 MLs, thus exhibiting a bell-shaped curve as a function of thickness (see Figure 5c). Note that a very similar dependence of the transport properties has been also observed for (001)-oriented LAO/STO interfaces. For instance, C. Bell et al. found that above the threshold thickness (> 4 MLs) the resistance of the (001)-interfaces increased steadily with thickness, while the mobility decreased steeply, being the effect more dramatic above 15 MLs [23].

High mobility conduction at (110) and (111) LaAlO$_3$/SrTiO$_3$ interfaces

**Discussion**

The results reported above reveal that (110), (100) and (111) LAO/STO interfaces, although differing on their polar nature, lead to similar conducting interfaces. This observation suggests that interface polarity may not be the crucial parameter for the observed behavior. To further cross check this view, we have grown on (110) STO substrates ultrathin oxides amorphous layers having cations with distinct oxygen affinity, namely: LAO (4.8 nm, equivalent to 18 (110) LAO MLs), STO (7.8 nm, equivalent to 29 (110) STO MLs) and YSZ (4.0 nm, equivalent to 10 (110) YSZ MLs). To obtain these amorphous overlayers, the growth has been performed at room temperature at the same pressure as that used for the growth of the LAO/STO interfaces described above ($P_{O2} = 10^{-4}$ mbar). The amorphous structure of these films was fully consistent with the observation of a diffuse halo in RHEED patterns and absence of film Bragg spots at any thickness during the growth (see Supplementary Figure S5). Remarkably, the amorphous (110) LAO/STO, STO/STO and YSZ/STO interfaces were found to be conductive, in agreement with recent reports on related (001) interfaces [25] and with values similar to those found for the epitaxial (110)- and (111)-oriented LAO/STO interfaces ($R_{xx} \approx 15 - 25$ k$\Omega$ at room temperature).

Obviously, a scenario based on polar discontinuity across the interface between bulk-like stacked atomic planes can not explain the observation of a metallic state in the (110) STO interface, either epitaxial or amorphous. Whatever the ultimate reason for the occurrence of such metallic behavior, it requires the involvement of phenomena not arising from pure electrostatic or electronic effects associated to ideal stacking of simply ionic (110) STO planes. We stress that it is assumed that the LAO growth occurs on (110) and (111) STO planes having atomic arrangements as in bulk STO. However, if

High mobility conduction at (110) and (111) LaAlO$_3$/SrTiO$_3$ interfaces

electronic and atomic reconstructions of the bare (110) and (111) STO surfaces occur, they may lead to significant charge redistributions that can radically change the surface polarization of these crystallographic planes as shown in Ref. [20], thus making rather insubstantial the differences among ideal (001), (110) and (111) STO surfaces. Still, the observation that low-temperature deposition of distinct amorphous oxides on (110) STO also promotes a metallic behavior is crucial and appears to resolve the *Gordian knot* of the interface metallic properties: while releasing the need of periodic atomic stacking and polar discontinuity, it may point to some cation intermixing or to oxygen depletion from the substrate during growth either related to highly energetic kinetic growth by pulsed laser deposition or/and the oxygen affinity of adatoms at the STO surface.

In short, we have shown that a metallic conductivity with low-temperature high carrier mobility can be induced in a variety of surfaces of SrTiO$_3$ single crystals, having distinct symmetry and polar nature, when interfaced with polar and non-polar oxides irrespectively on the crystalline or amorphous nature of the newly formed interfaces. Common to all reported results is that the 3$d^0$ empty orbitals of Ti ions play a fundamental role as electron acceptors, either resulting from electronic or chemical effects at interfaces with other oxides, or even with the vacuum [26]. As a natural extension of the present research and using similar approaches, the properties of electron gases along distinct crystalline planes may be analyzed in other $nd^0$ metal-oxide systems with more extended *d*-orbitals, such as KTaO$_3$, as a way to engineer novel interfaces with emerging properties.

High mobility conduction at (110) and (111) LaAlO$_3$/SrTiO$_3$ interfaces

## Methods

**Sample preparation.** Prior to film deposition, the STO substrates were treated in a dedicated furnace at 1100 ºC for 2 h under ambient conditions [27]. Subsequently, LAO thin films were grown by pulsed laser deposition ($\lambda$ = 248 nm) monitored by high pressure RHEED. The substrates were heated from room temperature to deposition temperature (850 ºC) in an oxygen partial pressure $P_{O2}$ = 0.1 mbar. Then, LAO was grown under $P_{O2}$ = 10$^{-4}$ mbar at 1 Hz repetition rate, with laser pulse energy of around 26 mJ. Two STO substrates, (110) and (111) oriented, were placed simultaneously for each deposition. The substrates were small in size (5 x 3 mm$^2$) to guarantee homogeneity. The in-situ RHEED analysis was done focusing the incident electrons on the STO(110) substrate and the LAO(110) thickness was controlled by monitoring the RHEED specular spot intensity (with incidence of electrons along STO[100] at a glancing angle of ~1º). Films with thickness 2, 4, 6, 7, 8, 10, 14, 18 and 36 LAO(110) MLs were prepared. It corresponds to a range of thickness from ~ 5 to ~ 100 Å, whereas the corresponding nominal range of (111) monolayers is ~ 2.5 to ~ 44. At the end of the deposition, samples were cooled down in oxygen rich atmosphere to minimize the formation of oxygen vacancies [3, 19]: $P_{O2}$ = 0.3 mbar from T = 850 ºC to 750 ºC and $P_{O2}$ = 200 mbar from T = 750 ºC to room temperature, with a dwell time of 1 hour at 600 ºC. Also, two (110) and (111) oriented substrates were exposed to the conditions of the growth process, being maintained at 850 ºC under 10$^{-4}$ mbar for 1000 s, but without LAO deposition. These samples are labelled here as 0 MLs films. Ultrathin amorphous films of LAO, STO, and YSZ, with nominal thicknesses of 4.8, 7.8 and 4.0 nm, respectively, were also deposited under 10$^{-4}$ mbar of oxygen.

High mobility conduction at (110) and (111) LaAlO$_3$/SrTiO$_3$ interfaces

**Sample characterization.** The surface morphology was analyzed by AFM working in dynamic mode (images were analyzed using the WSxM software [28]). The electrical characterization of the samples was performed by using six-contact arrangement in Hall geometry, from which the sheet resistance ($R_{xx}$), sheet carrier density ($n_{sheet}$) and electron mobility ($\mu$) were extracted as a function of temperature. The current was injected along the in-plane [001] direction in (110)-interfaces, whereas it was injected along [1 1 -2] in (111)-interfaces. The transverse resistance ($R_{xy}$) was linear for all the conductive samples and within all the range of applied magnetic fields up to ±9 T and temperatures (5 K ≤ T ≤ 300 K), and its field-dependence indicated *n*-type conduction, as found in (001) LAO/STO interfaces [1,2]. The LAO/STO interface was contacted via ultrasonic wire bonder with Al wires.

High mobility conduction at (110) and (111) LaAlO$_3$/SrTiO$_3$ interfaces

**Acknowledgements.**

Financial support by the Spanish Government [Projects MAT2011-29269-CO3 and NANOSELECT CSD2007-00041] and Generalitat de Catalunya (2009 SGR 00376) is acknowledged. We also acknowledge technical support from E. León from the ICMAB-CSIC and M. Stengel for careful reading of the manuscript and comments.

**Author contributions.**

F.S., J.F. and G.H. designed the experiments. M.S. performed atomic force microscopy analysis. M.S., F.S. and N.D. were responsible for thin film growth and RHEED analysis. G.H. carried out the magnetotransport experiments and analysis. G.H., F.S. and J.F. wrote the manuscript.

**Additional information**

**Supplementary information** accompanies this paper at http://www.nature.com/scientificreports

**Competing financial interests**: The authors declare no competing financial interests.

High mobility conduction at (110) and (111) LaAlO$_3$/SrTiO$_3$ interfaces

**Figure 1. Highly conductive interfaces.** The sketches display (001)-, (110)- and (111)-oriented LAO/STO interfaces, along with oxide amorphous layers (LAO, STO, YSZ) interfacing (110)-oriented STO, all of them exhibiting high-mobility conduction.

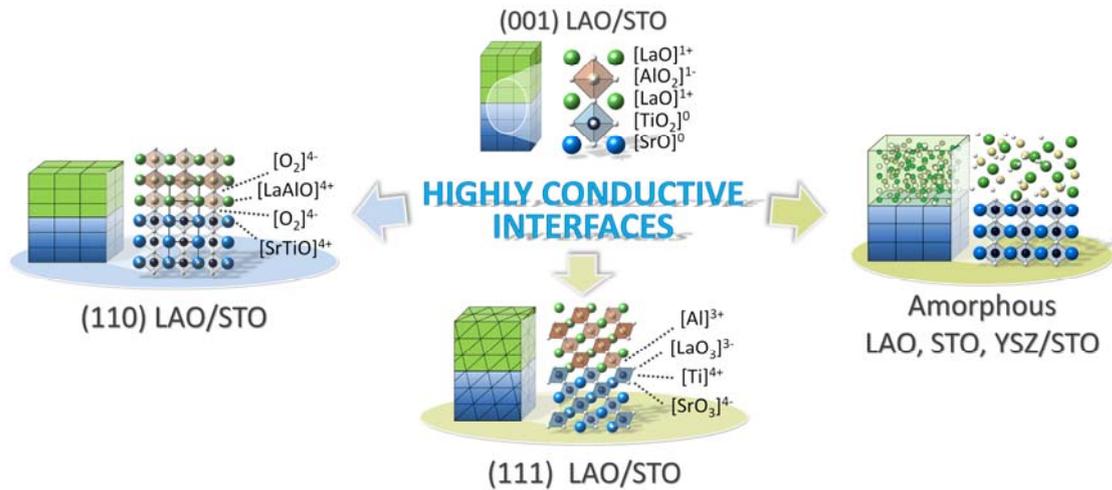

High mobility conduction at (110) and (111) $LaAlO_3/SrTiO_3$ interfaces

**Figure 2. RHEED and AFM.** (**a**) RHEED pattern taken along [001] at 850 ºC and $10^{-4}$ mbar of the (110) STO substrate used to grow the 36 MLs (110) LAO sample. The pattern at the end of growth ($P_{O2}=1\times10^{-4}$ mbar) is shown in (**c**). (**b**) Representative curves of the intensity of the specular spot for several (110) LAO films of thickness: 4 MLs (bottom), 10 MLs (middle), and 36 MLs (up). In the latter case, only the intensity corresponding to the first 14 MLs is shown. The topographic AFM images corresponding to these films are in (**d-f**), respectively.

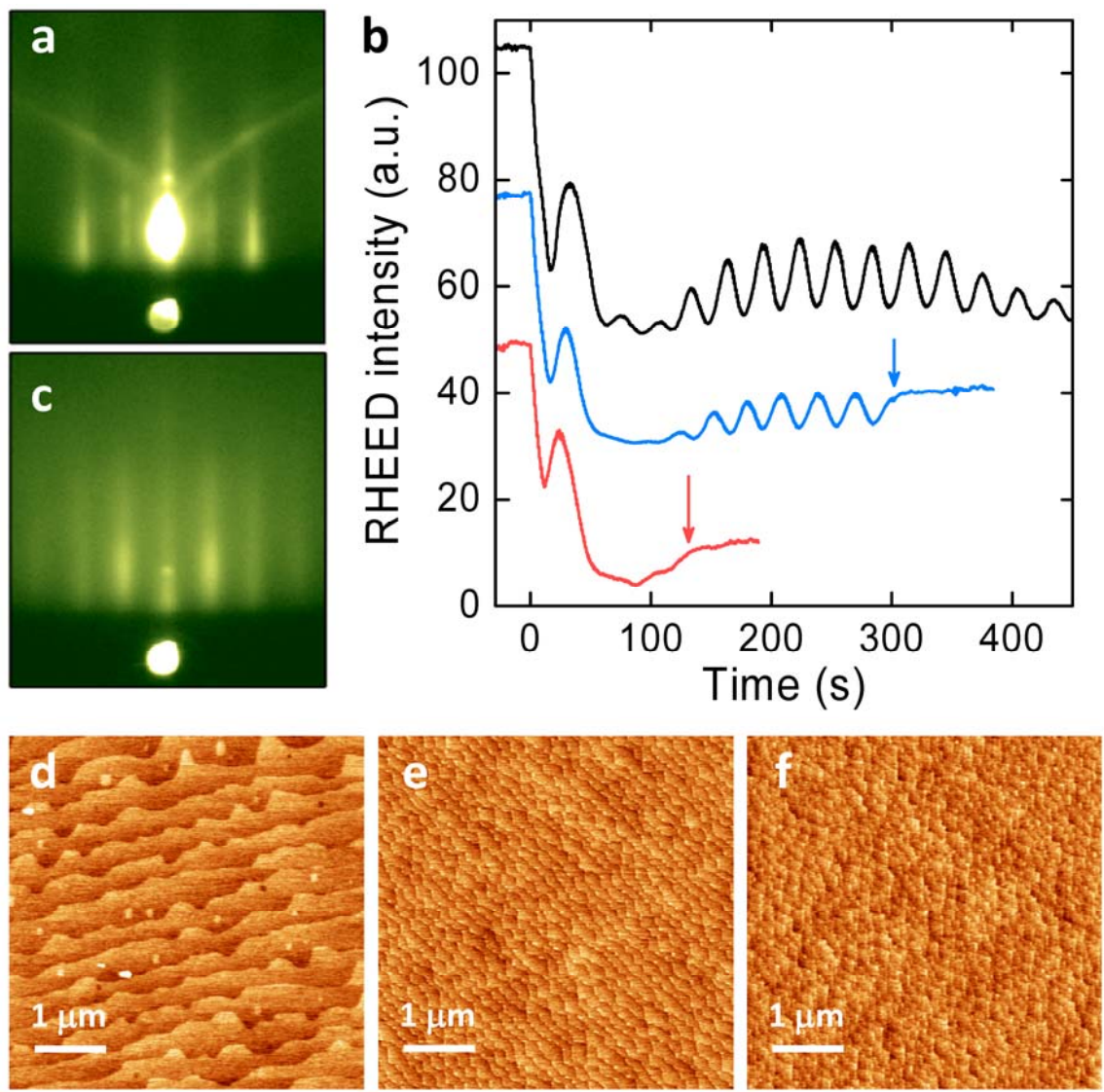

High mobility conduction at (110) and (111) LaAlO$_3$/SrTiO$_3$ interfaces

**Figure 3. Critical thickness for conductive LAO/STO interfaces.** Conductance measured at room temperature as a function of the LAO overlayer thickness for interfaces oriented along [110], [001], and [111] directions. The data corresponding to the (001)-oriented interfaces are collected from References [19] and [23].

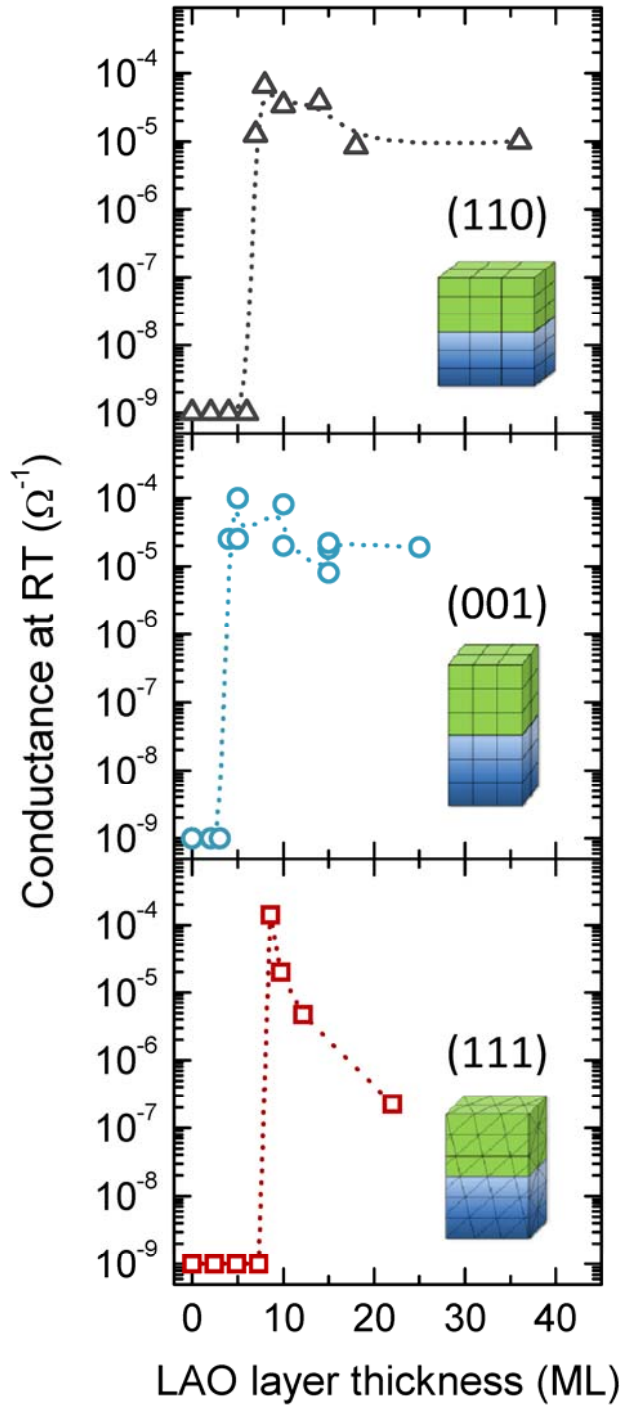

High mobility conduction at (110) and (111) LaAlO$_3$/SrTiO$_3$ interfaces

**Figure 4. Sheet resistance of LAO/STO interfaces.** Temperature dependence of the sheet resistance of LAO/STO interfaces of different LAO overlayer thickness, oriented along (**a**) [110]- and (**b**) [111]- crystallographic directions.

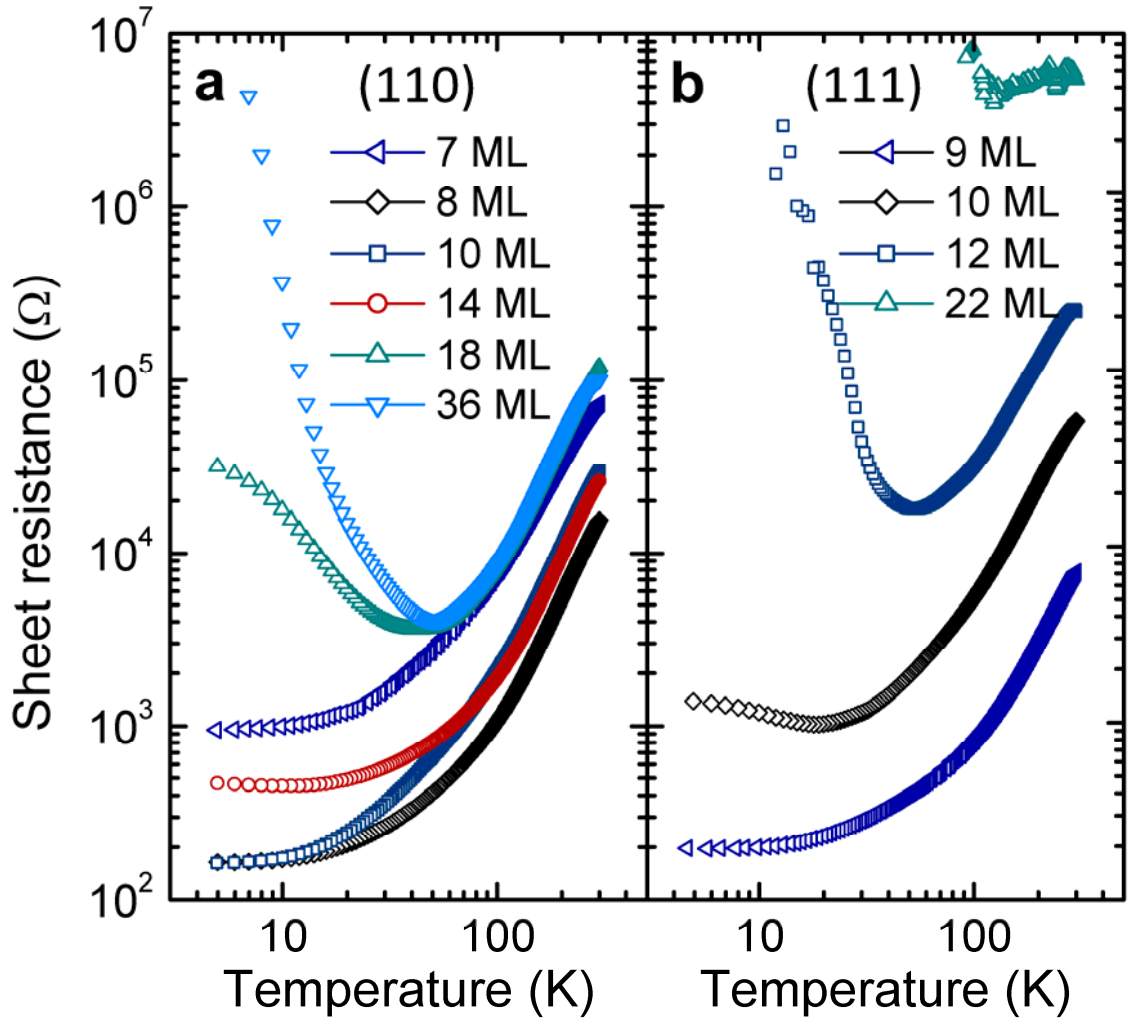

High mobility conduction at (110) and (111) LaAlO$_3$/SrTiO$_3$ interfaces

**Figure 5. Sheet carrier density and mobility.** Temperature dependence of the (**a**) sheet carrier density and (**b**) mobility of (110)-oriented LAO/STO interfaces with different LAO overlayer thickness. (**c**) The dependence of the mobility at T = 5 K is also plotted as a function of the measured sheet carrier density at each value of the LAO overlayer thickness. The dotted line is a guide to the eye.

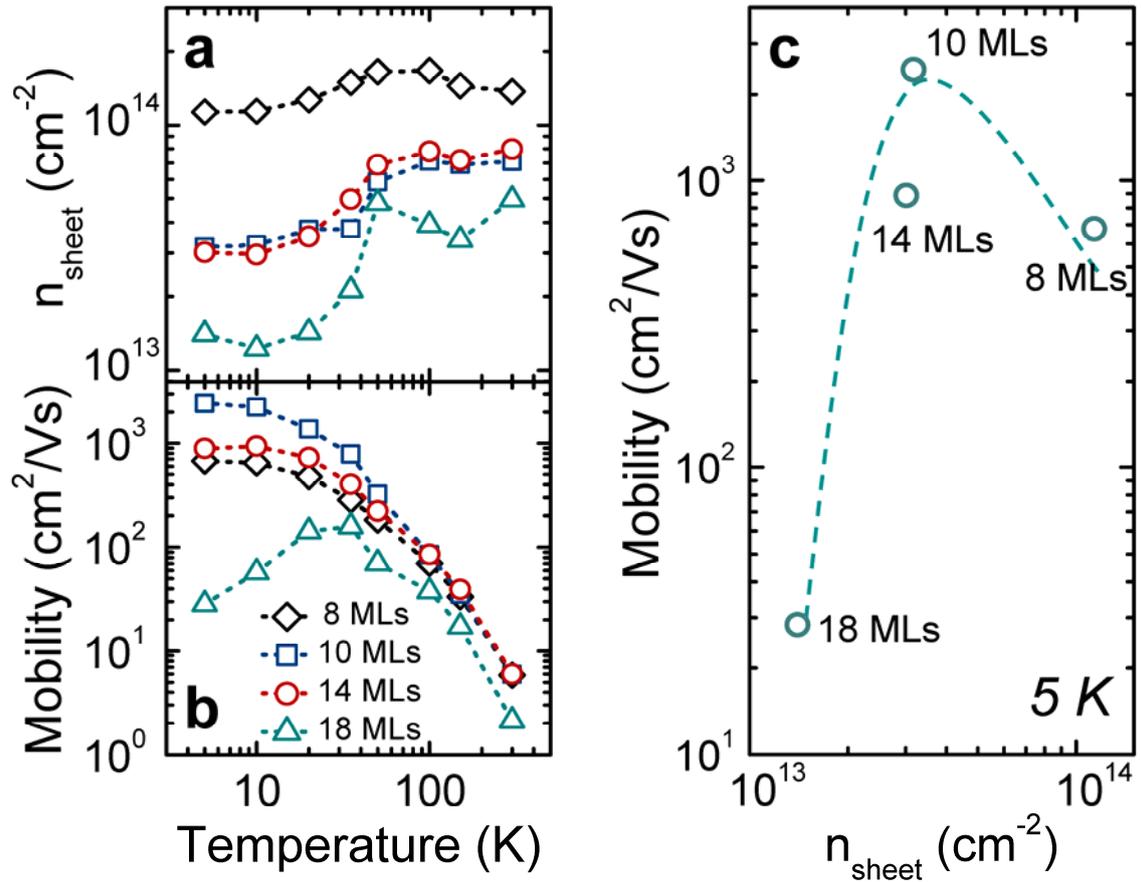

# Supplementary Information: High mobility conduction at (110) and (111) LaAlO$_3$/SrTiO$_3$ interfaces

G. Herranz, F. Sánchez, N. Dix, M. Scigaj, J. Fontcuberta

Institut de Ciència de Materials de Barcelona (ICMAB-CSIC), Campus de la UAB, Bellaterra 08193, Catalonia, Spain

**Supplementary Table S1**.

| ABO$_3$ | (001) | (110) | (111) |
|---|---|---|---|
| | AO<br>BO$_2$ | ABO<br>O$_2$ | AO$_3$<br>B |
| **LaAlO$_3$ (LAO)**<br>**SrTiO$_3$ (STO)** | [1 0 0]  [0 1 0]<br>[AlO$_2$]$^{1-}$ / [LaO]$^{1+}$ / [TiO$_2$]$^0$ / [SrO]$^0$ | [1 -1 0]  [0 0 1]<br>[O$_2$]$^{4-}$ / [LaAlO]$^{4+}$ / [O$_2$]$^{4-}$ / [SrTiO]$^{4+}$ | [-1 1 0]  [1 1 -2]<br>[Al]$^{3+}$ / [LaO$_3$]$^{3-}$ / [Ti]$^{4+}$ / [SrO$_3$]$^{4-}$ |
| **polarity discontinuity** | YES | NO | YES |

# Supplementary Information: High mobility conduction at (110) and (111) LaAlO$_3$/SrTiO$_3$ interfaces

**Supplementary Figure S1**: AFM topographic images of (110) SrTiO$_3$ (a) and (111) SrTiO$_3$ (b) substrates treated 2h at 1100 ºC. Closer views are shown in (c) and (e). The corresponding height profiles along the lines are shown in panels (g) and (h). RHEED patterns acquired at room temperature and high vacuum taken along [001] and [11-2] are shown in (d) and (f), respectively. High-quality surfaces are revealed by the presence of Bragg spots along the 0$^{th}$ and the 1$^{st}$ Laue circles and Kikuchi lines.

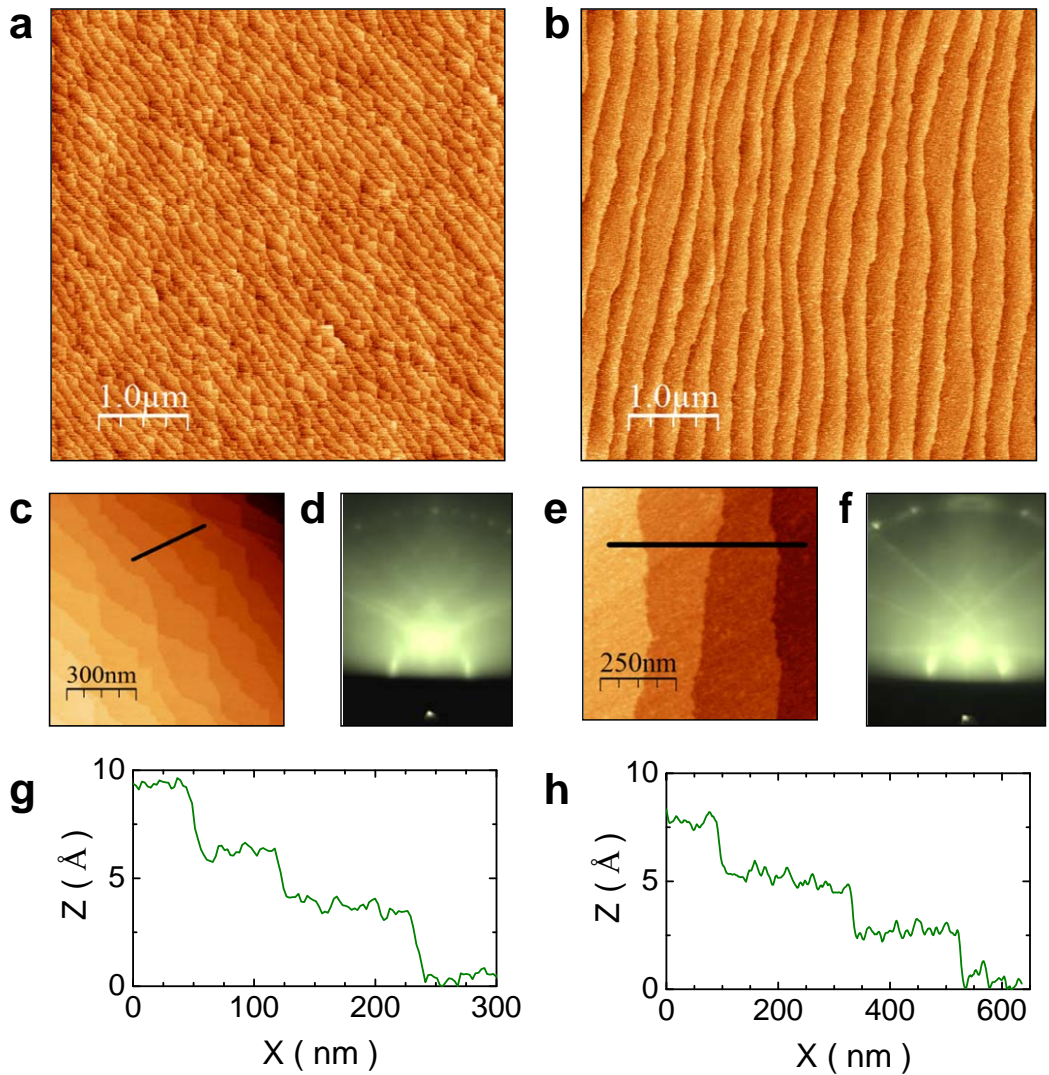

# Supplementary Information: High mobility conduction at (110) and (111) LaAlO$_3$/SrTiO$_3$ interfaces

**Supplementary Figure S2**: X-Ray reflectivity data of the 36 MLs (110) LaAlO$_3$/SrTiO$_3$ sample (black curve). The simulation (red curve) corresponds to a LaAlO$_3$ thickness of 96.3 Å, very close to the thickness of 36 MLs of relaxed (110) LaAlO$_3$ (96.5 Å).

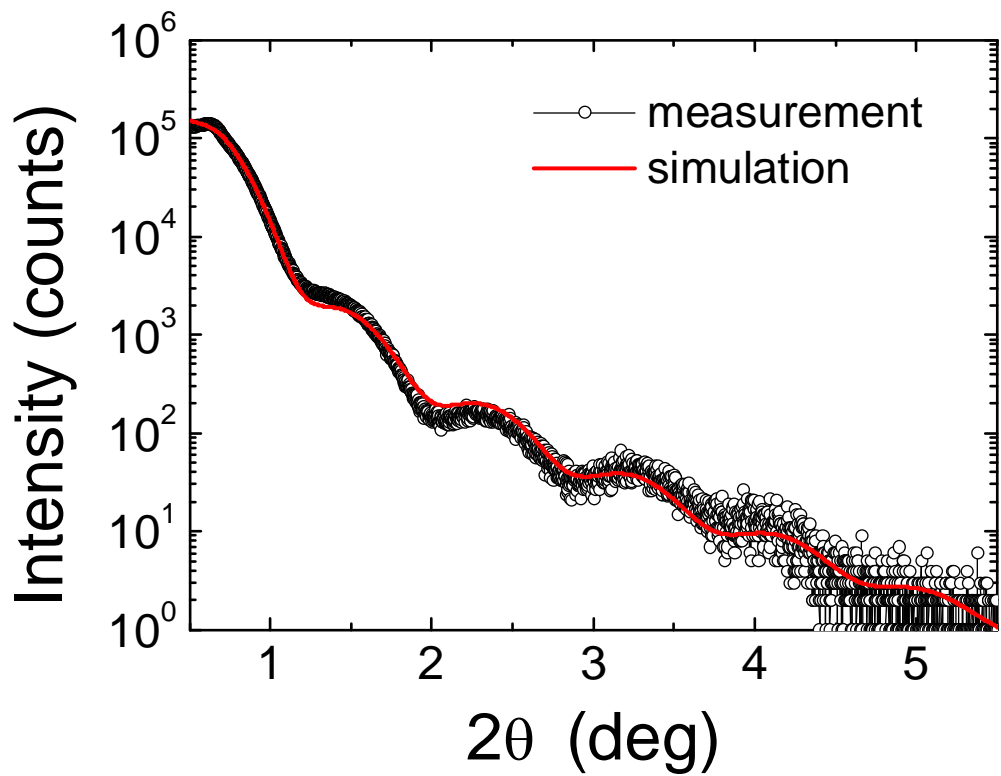

# Supplementary Information: High mobility conduction at (110) and (111) LaAlO$_3$/SrTiO$_3$ interfaces

**Supplementary Figure S3**: Atomic force microscopy topographic image of the 9 (111) LaAlO$_3$ MLs on SrTiO$_3$ sample (a) and the RHEED pattern taken along the [11-2] direction (b).

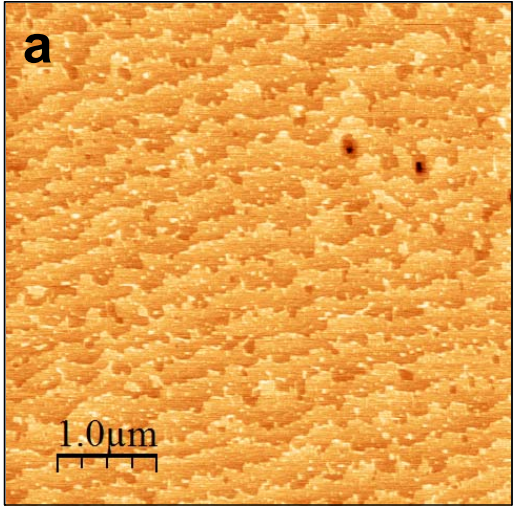
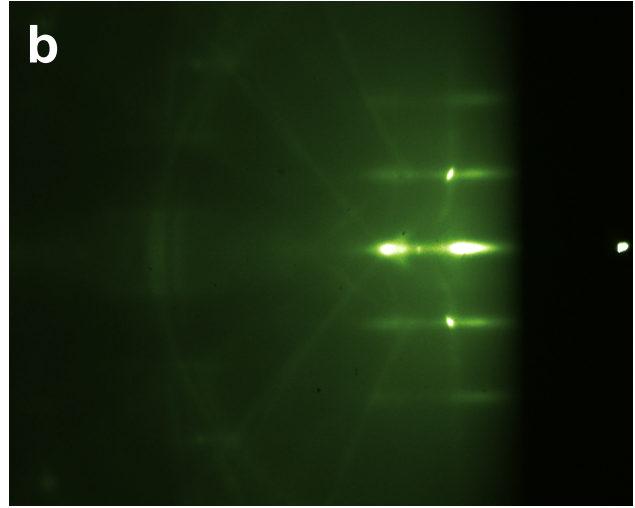

# Supplementary Information: High mobility conduction at (110) and (111) LaAlO$_3$/SrTiO$_3$ interfaces

**Supplementary Figure S4**: Temperature dependence of the resistance of 10 MLs of LAO deposited on two different STO substrates. A first substrate was treated at 1100 ºC for 2 h under ambient conditions before growing the LAO overlayer (sample A), as done for all films reported in the manuscript (see Methods). A second LAO film (sample B) was grown on a substrate that was annealed in-situ at 900 ºC under base pressure (~10$^{-7}$ mbar) for 1h and subsequently cooled down to 550ºC and maintained at that temperature for 1 hour under a pressure of 1x10$^{-4}$ Torr.

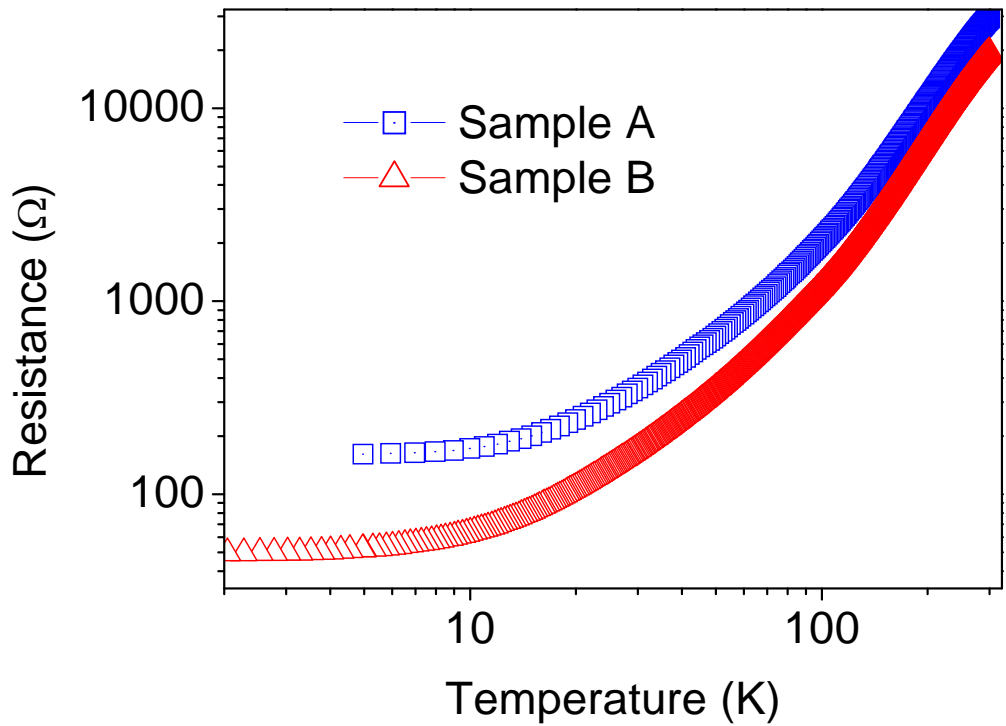

# Supplementary Information: High mobility conduction at (110) and (111) LaAlO$_3$/SrTiO$_3$ interfaces

**Supplementary Figure S5**: RHEED patterns acquired at a grazing angle ~0.9º for films deposited at room temperature on STO(110): ~4.8 nm LAO (a), ~4.0 nm YSZ (b) and ~7.8 nm STO (c). The absence of Bragg spots and the observation of a diffusion halo are consistent with the amorphous structure of these films. The bright arc observed in the figures is an instrumental artifact that remains in the same screen position (not shown here) if the film is removed.

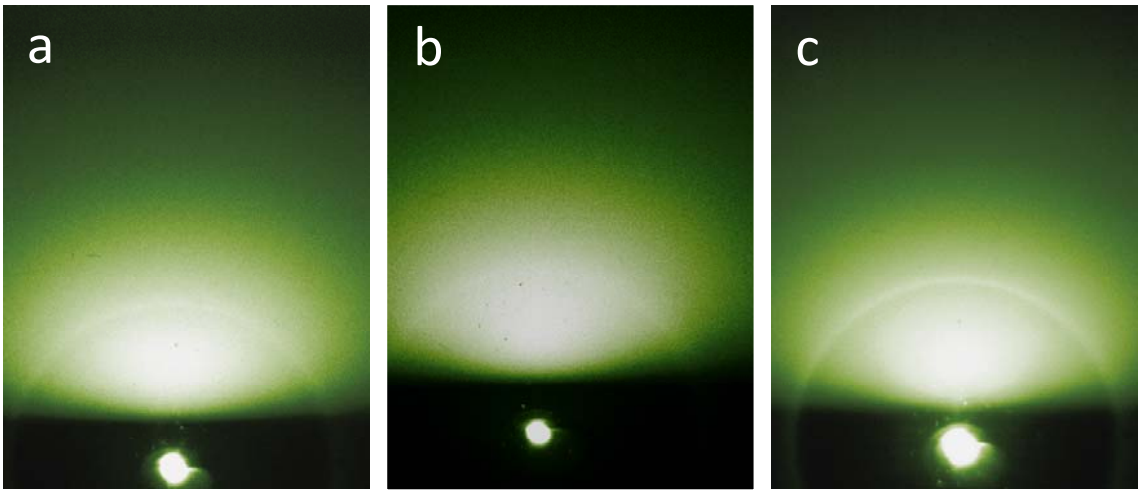